\documentclass{revtex4}
\usepackage{amsmath}

\input{tcilatex}

\begin{document}

\title[ ]{Domain-Walls in Einstein-Gauss-Bonnet Bulk}
\author{S. Habib Mazharimousavi}
\email{habib.mazhari@emu.edu.tr}
\author{M. Halilsoy}
\email{mustafa.halilsoy@emu.edu.tr}
\affiliation{Department of Physics, Eastern Mediterranean University, G. Magusa, north
Cyprus, Mersin 10, Turkey. }

\begin{abstract}
We investigate the dynamics of a d-dimensional domain wall (DW) in a
d+1-dimensional Einstein-Gauss-Bonnet (EGB) bulk. Exact effective potential
induced by the Gauss-Bonnet (GB) term on the wall is derived. In the absence
of the GB term we recover the familiar gravitational and anti-harmonic
oscillator potentials. Inclusion of the GB correction gives rise to a
minimum radius of bounce for the Friedmann-Robertson-Walker (FRW) universe
expanding with a negative pressure on the DW.
\end{abstract}

\maketitle

We consider a $n-$dimensional domain wall (DW) $\Sigma $ in a $n+1-$%
dimensional bulk $\mathcal{M}$. This DW splits the background bulk into two $%
n+1-$dimensional spacetimes which will be referred to as $\mathcal{M}_{\pm
}. $ Here $\pm $ is assumed with respect to the DW. Our action of
Gauss-Bonnet (GB) extended gravity is chosen as 
\begin{equation}
S=\frac{1}{2\kappa ^{2}}\int_{\mathcal{M}}d^{n+1}x\sqrt{-g}\left( R+\alpha 
\mathcal{L}_{GB}\right) +\frac{1}{\kappa ^{2}}\int_{\Sigma }d^{n}x\sqrt{-h}%
\left\{ K\right\} +\int_{\Sigma }d^{n}x\sqrt{-h}\mathcal{L}_{DW},
\end{equation}%
in which $\mathcal{L}_{DW}=-\sigma =$constant is the Nambu-Goto form of DW
Lagrangian, and $K$ is the extrinsic curvature of DW with $h=$ $\left\vert
g_{ij}\right\vert .$ (Latin indices run over the DW coordinates while Greek
indices refer to the bulk's coordinates). The GB Lagrangian $\mathcal{L}%
_{GB} $ is given by 
\begin{equation}
\mathcal{L}_{GB}=R_{\mu \nu \gamma \delta }R^{\mu \nu \gamma \delta
}-4R_{\mu \nu }R^{\mu \nu }+R^{2},
\end{equation}%
with the GB parameter $\alpha .$ Variation of the action with respect to the
space-time metric $g_{\mu \nu }$ yields the field equations%
\begin{equation}
G_{\mu \nu }^{E}+\alpha G_{\mu \nu }^{GB}=0,
\end{equation}%
where 
\begin{equation}
G_{\mu \nu }^{GB}=2\left( -R_{\mu \sigma \kappa \tau }R_{\quad \nu }^{\kappa
\tau \sigma }-2R_{\mu \rho \nu \sigma }R^{\rho \sigma }-2R_{\mu \sigma }R_{\
\nu }^{\sigma }+RR_{\mu \nu }\right) -\frac{1}{2}\mathcal{L}_{GB}g_{\mu \nu }%
\text{ .}
\end{equation}%
Our bulk metric is a $n+1-$dimensional static, spherically symmetric
spacetime, 
\begin{equation}
ds_{b}^{2}=-f\left( r\right) dt^{2}+\frac{1}{f\left( r\right) }%
dr^{2}+r^{2}d\Omega _{n-1}^{2}
\end{equation}%
in which $f(r)$ is the only metric function to be determined and $d\Omega
_{n-1}^{2}$ is the line element of $S^{n-1}.\ $Upon imposing the constraint 
\begin{equation}
-f\left( a\right) \left( \frac{dt}{d\tau }\right) ^{2}+\frac{1}{f\left(
a\right) }\left( \frac{da}{d\tau }\right) ^{2}=-1
\end{equation}%
with the DW position at $r=a\left( \tau \right) ,$ the DW's line element
takes the form 
\begin{equation}
ds_{dw}^{2}=-d\tau ^{2}+a\left( \tau \right) ^{2}d\Omega _{n-1}^{2}.
\end{equation}%
This is the standard FRW metric and its only degree of freedom is $a\left(
\tau \right) $ in which $\tau $ is the proper time measured by the observer
on the DW. Now, we wish to consider the rules satisfied by the DW as the
boundary of $\mathcal{M}_{\pm }.$ These boundary conditions are the
generalized Israel conditions which correspond to the Einstein equations on
the wall. \cite{1}

The generalized Darmois-Israel junction conditions on $\Sigma $ \ apt for
the GB extension is \cite{2}%
\begin{equation}
-\frac{1}{\kappa ^{2}}\left( \left\langle K_{i}^{j}\right\rangle -K\delta
_{i}^{j}\right) -\frac{\alpha }{2\kappa ^{2}}\left\langle 3J_{i}^{j}-J\delta
_{i}^{j}+2P_{imn}^{\;\;\;\;j}K^{mn}\right\rangle =S_{i}^{j},
\end{equation}%
where the surface energy-momentum tensor $S_{ij}$ is given by \cite{3}%
\begin{equation}
S_{ij}=\frac{1}{\sqrt{-h}}\frac{2\delta }{\delta g^{ij}}\int d^{n}x\sqrt{-h}%
\left( -\sigma \right) .
\end{equation}%
The form of the stress-energy tensor can be written as%
\begin{equation}
S_{i}^{j}=-\sigma \delta _{i}^{j}
\end{equation}%
in which $\sigma =$constant, stands for the wall tension (or energy density
of the wall $\Sigma $). Considering the energy-momentum tensor in the form $%
S_{i}^{j}=$diag$\left( -\rho ,p,p,...\right) ,$ we observe that $\sigma
=\rho =-p$, and satisfies the Weak Energy Condition. Here in (8) a bracket
implies a jump across $\Sigma .$ The divergence-free part of the Riemann
tensor $P_{abcd}$ and the tensor $J_{ab}$ (with trace $J=J_{a}^{a}$) are
given by \cite{2} 
\begin{align}
P_{imnj}& =R_{imnj}+\left( R_{mn}g_{ij}-R_{mj}g_{in}\right) -\left(
R_{in}g_{mj}-R_{ij}g_{mn}\right) +\frac{1}{2}R\left(
g_{in}g_{mj}-g_{ij}g_{mn}\right) , \\
J_{ij}& =\frac{1}{3}\left[
2KK_{im}K_{j}^{m}+K_{mn}K^{mn}K_{ij}-2K_{im}K^{mn}K_{nj}-K^{2}K_{ij}\right] .
\end{align}%
By employing these expressions through (8) and (10) we find the energy
density and surface pressures for a generic metric function $f\left(
r\right) ,$ with $r=a\left( \tau \right) .$ The results are given by \cite{4}
\begin{gather}
-\Delta \left( n-1\right) \left[ \frac{2}{a}-\frac{4\tilde{\alpha}}{3a^{3}}%
\left( \Delta ^{2}-3\left( 1+\dot{a}^{2}\right) \right) \right] =\kappa
^{2}\sigma , \\
\begin{tabular}{l}
$\frac{2\left( n-2\right) \Delta }{a}+\frac{2\ell }{\Delta }-\frac{4\tilde{%
\alpha}}{3a^{2}}\left[ 3\ell \Delta -\frac{3\ell }{\Delta }\left( 1+\dot{a}%
^{2}\right) +\frac{\Delta ^{3}}{a}\left( n-4\right) -\frac{6\Delta }{a}%
\left( a\ddot{a}+\frac{n-4}{2}\left( 1+\dot{a}^{2}\right) \right) \right]
=-\kappa ^{2}\sigma $%
\end{tabular}%
,
\end{gather}%
where $\ell =\ddot{a}+f^{\prime }\left( a\right) /2$ and $\Delta =\sqrt{%
f\left( a\right) +\dot{a}^{2}}$ in which 
\begin{equation}
f\left( a\right) =\left. f_{-}\left( r\right) \right\vert _{r=a}.
\end{equation}%
Note that a dot "$\cdot $" implies derivative with respect to the proper
time.

We differentiate (13) to get (with $\ddot{a}=\ell -f^{\prime }\left(
a\right) /2$) 
\begin{equation}
\ell =\frac{\Delta ^{2}}{a}\frac{a^{2}+\tilde{\alpha}\left[ 2\left(
af^{\prime }-\Delta ^{2}\right) +6\left( 1+\dot{a}^{2}\right) \right] }{%
a^{2}+2\tilde{\alpha}\left( \Delta ^{2}+1+\dot{a}^{2}\right) },
\end{equation}%
which, after substitution into (14) we recover (13). In other words, Eq.s
(13) and (14) are not independent, the solution of one satisfies also the
other. Now, we analyze the first equation (13) of the junction conditions.
By some manipulation, $\Delta $ above can be expressed in the form%
\begin{equation}
\Delta =\sqrt[3]{\xi }-\frac{\epsilon }{3\sqrt[3]{\xi }},
\end{equation}%
where 
\begin{eqnarray}
\xi &=&-\frac{1}{2}s\pm \frac{1}{18}\sqrt{12\epsilon ^{3}+81s^{2}} \\
s &=&\frac{3}{8}\frac{\kappa ^{2}\sigma a^{3}}{\tilde{\alpha}\left(
d-1\right) },\epsilon =\frac{3}{2}\left( 1-f+\frac{a^{2}}{2\tilde{\alpha}}%
\right) .
\end{eqnarray}%
From $\Delta =\sqrt{f\left( a\right) +\dot{a}^{2}}$ and (17) it follows that%
\begin{equation}
\dot{a}^{2}+V\left( a\right) =0
\end{equation}%
where%
\begin{equation}
V\left( a\right) =f-\left( \sqrt[3]{\xi }-\frac{\epsilon }{3\sqrt[3]{\xi }}%
\right) ^{2}.
\end{equation}

In the sequel we consider the wall to be a classical one-dimensional
particle which moves with zero total energy under the effective potential $%
V\left( a\right) .$ It is clear from (20) that only $V(a)<0$ has a physical
meaning. By plotting $V\left( a\right) $ in terms of $a$ we investigate the
possible types of motion for the wall.

The metric function $f\left( r\right) $ is the solution of the Einstein
equations in the $n+1-$dimensional bulk, i.e. from Eq. (5). In terms of the
ADM mass and GB parameter $\tilde{\alpha}=\left( n-2\right) \left(
n-3\right) \alpha ,$ solution for $f\left( r\right) $ is \cite{5} 
\begin{equation}
f_{\pm }\left( r\right) =1+\frac{r^{2}}{2\tilde{\alpha}}\left( 1\pm \sqrt{1+%
\frac{16\tilde{\alpha}M_{ADM}}{\left( n-1\right) r^{n}}}\right) .
\end{equation}%
Here, the negative branch gives the correct limit of general relativity,
i.e., 
\begin{eqnarray}
\lim_{\tilde{\alpha}\rightarrow 0}f_{-}\left( r\right) &=&1-\frac{4M_{ADM}}{%
\left( n-1\right) r^{n-2}}, \\
\lim_{\tilde{\alpha}\rightarrow \infty }f_{-}\left( r\right) &=&1.
\end{eqnarray}%
For this reason we consider the negative branch solution, which means that $%
f\left( a\right) =f_{-}\left( a\right) .$ Upon substitution of $f\left(
a\right) $ in (21) we observe that%
\begin{equation}
\lim_{\tilde{\alpha}\rightarrow \infty }V\left( a\right) =1,
\end{equation}%
which corresponds to a non-physical case (i.e. Eq. (20)) and%
\begin{equation}
\lim_{\tilde{\alpha}\rightarrow 0}V\left( a\right) =V_{0}=1-\frac{4M_{ADM}}{%
\left( n-1\right) a^{n-2}}-\frac{1}{4}\frac{\kappa ^{4}a^{2}\sigma ^{2}}{%
\left( n-1\right) ^{2}}.
\end{equation}%
This shows that vanishing of the GB parameter yields a potential on the DW
which contains a gravitational and anti-harmonic oscillator potentials. The
exact potential (with $\alpha \neq 0$), however, has a rather intricate
structure which can be expanded in terms of the $\alpha $ as%
\begin{equation}
V\left( a\right) =V_{0}+V_{1}\alpha +V_{2}\alpha ^{2}+...
\end{equation}%
for $V_{0}$ was given in Eq. (26) 
\begin{equation}
V_{1}=\frac{\left( n-2\right) \left( n-3\right) }{\left( n-1\right) ^{2}}%
\left( \frac{\sigma ^{4}\kappa ^{8}a^{2}}{6\left( n-1\right) ^{2}}+\frac{%
4M\sigma ^{2}\kappa ^{4}}{\left( n-1\right) a^{n-2}}+\frac{16\left(
n-3\right) M^{2}}{a^{2\left( n-1\right) }}\right) ,
\end{equation}%
and 
\begin{equation}
V_{2}=\frac{\left( n-2\right) ^{2}\left( n-3\right) ^{2}}{\left( n-1\right)
^{3}}\left( -\frac{7}{36}\frac{\sigma ^{6}\kappa ^{12}a^{2}}{\left(
n-1\right) ^{3}}-\frac{20}{3}\frac{M\sigma ^{4}\kappa ^{8}}{\left(
n-1\right) ^{2}a^{n-2}}-\frac{64M^{2}\sigma ^{2}\kappa ^{4}}{\left(
n-1\right) a^{2\left( n-1\right) }}-\frac{128M^{3}}{a^{3n-25}}\right) .
\end{equation}

In Fig.s (1-3) we display $V(a)$ and $f(a)$ for $\kappa ^{2}=1,\sigma =1,$ $%
n=4$, with changing $\alpha $ and $M$. For different $\alpha $ and $M$
values we may obtain similar plots, such as for example $2a$ and $3c$. This
implies that the effect of $\alpha $ may be compensated with that of $M$ and
vice versa. Once inside the event horizon of the black hole the DW has no
chance but crush to the central singularity as it should. This is the
ultimate fate of our DW universe if it lies inside a large black hole. For
favorable condition of the potential (i.e. $V(a)<0$) and in the vicinity
(outside) of the horizon the DW collapses into the black hole much like
shells \cite{6}. The overall view, however, whether we have a black hole or
not is that the potential provides a minimum bounce for the DW which is
determined by the GB parameter $\alpha .$

We should also add that in our analysis we were unable to see a maximum
bounce. This implies that the GB extension of general relativity doesn't
suffice to provide a closed universe on DW.

Fig. 4 plots the same quantities in $n=5$, for comparison with the previous
ones in $n=4$. What we observe is that going into higher dimensions does not
change the general features except that some non-black hole cases will turn
into black holes as well. We should remark also that although the coupling
constant $\sigma $ between the bulk and DW has been fixed as $\sigma =1$,
its effect can be investigated by taking different values for $\sigma .$ In
general, larger $\sigma $ results smaller bouncing radii and vice versa.

In conclusion, if our $4-$dimensional universe, assumed as a FRW universe on
a DW laying in a $5-$dimensional Einstein-Gauss-Bonnet (EGB) bulk, the GB
term protects us against the big-crunch. Inclusion of physical fields such
as Maxwell and Yang-Mills will definitely enrich our world on such a DW.
Abiding by a bulk consisting of pure geometrical terms alone, however, the
hierarchy of GB, known as the Lovelock gravity must be taken into account.

\bigskip

\bigskip

\textbf{Figure Caption:}

Fig. 1: For $5-$dimensional bulk, we have our DW as $4-$dimensional FRW
universe. With increasing $\alpha $ the bouncing radius of the DW universe
increases also. Fig. 1a is a black hole, while Fig. 1b can be interpreted as
a pointlike black hole. Fig.s 1c and 1d are non-black hole cases with
differing bounce radii. The arrows show the possible motions of DW including
bounces.

Fig. 2: Beside the minimum bouncing radii in the DW the smaller region
between horizon and allowable potential may have a DW which has no chance
other than collapsing into the black hole. Fig. 2a has a critical radius $%
a_{c}$ for which $V\left( a_{c}\right) =0.$ The nature of a DW inside the
black hole of course changes, since it turns into a dynamic and collapsing
object toward the central singularity. This occurs in 2a and 2b more
clearly. Fig. 2d is similar to 1b, which means that the mass difference
compensates with the difference in $\alpha .$

Fig. 3: The minimum bounces of the DW universe in 3a and 3b occur at the
horizon so that the DW collapses into the black hole. In 3c we have also $%
V\left( a_{c}\right) =0$. For $a>a_{c}$, the bounce does not occur at $a_{c}$%
. For $a<a_{c}$ the DW collapses into the black hole while in 3d, it has no
chance to fall into the black hole. Once inside the horizon, its fate ends
at the central singularity.

Fig. 4: For $6-$dimensional bulk, we have DW as a $5-$dimensional universe.
Fig 4a / 4b has similar feature with 3c / 3d. Fig 4c is also similar to 2b.
Fig. 4d represents a black hole with very small horizon but with very large
bouncing radius.

\bigskip

\end{document}